\documentclass[11pt]{article}
\usepackage{rotating}

\usepackage{graphicx}
\begin{document}
\title{\bf Peculiarities of $\alpha$-element Abundances in Galactic 
Open Clusters}

\author{{V.\,A.~Marsakov, M.\,L.Gozha, V.\,V.~Koval', L.\,V. Shpigel'}\\
{Southern Federal University, Rostov-on-Don, Russia}\\
{e-mail:  marsakov@sfedu.ru, vvkoval@sfedu.ru}}
\date{accepted \ 2016, Astronomy Reports, Vol. 60, No. 1, pp. 43-60}

\maketitle

\begin {abstract}

A catalog compiling the parameters of 346 open clusters, including 
their metallicities, positions, ages, and velocities has been composed. 
The elements of the Galactic orbits for 272 of the clusters
have been calculated. Spectroscopic determinations of the relative 
abundances, [el/Fe], for 14 elements synthesized in various nuclear 
processes averaged over data from 109 publications are presented for
90 clusters. The compiled data indicate that the relative 
abundances of primary $\alpha$\,-elements (oxygen and magnesium) exhibit
different dependences on metallicity, age, Galactocentric distance, and 
the elements of the Galactic orbits in clusters with high, elongated 
orbits satisfying the criterion $(Z^{2}_{max} + 4e^{2})^{1/2} > 0.40$
and in field stars of the Galactic thin disk ($Z_{max}$ is the maximum 
distance of the orbit from the Galactic plane in kiloparsec and $e$ is 
the eccentricity of the Galactic orbit). Since no systematic effects 
distorting the relative abundances of the studied elements in these 
clusters have been found, these difference suggest real differences 
between clusters with high, elongated orbits and field stars. 
In particular, this supports the earlier conclusion, based on an 
analysis of the elements of the Galactic orbits, that some clusters formed
as a result of interactions between high-velocity, metal-poor clouds 
and the interstellar medium of the Galactic thin disk. On average, 
clusterswith high, elongated orbits and metallicities 
${\rm [Fe/H]} < - 0.1$ 
display lower relative abundances of the primary б elements than do 
field stars. The low [O,Mg/Fe] ratios of these clusters can be 
understood if the high-velocity clouds that gave rise to them were 
formed of interstellar material from regions where the star-formation 
rate and/or the masses of Type II supernovae were lower than near the 
Galactic plane. It is also shown that, on average, the relative 
abundances of the primary б elements are higher in relatively 
metal-rich clusters with high, elongated orbits than in field stars. 
This can be understood if clusters with ${\rm [Fe/H]} > - 0.1$ 
formed as a 
result of interactions between metal-rich clouds with intermediate 
velocities and the interstellar medium of the Galactic disk; such 
clouds could form from returning gas in a so-called ``Galactic fountain''.

\end{abstract}

{{\bf Key words:} open star clusters, chemical composition, kinematics, 
Galaxy (Milky Way)}.

\maketitle

\section{Introduction}

The relative elemental abundances [el/Fe] in stellar
atmospheres carry information about the chemical
evolution of the interstellar material from which the
stars were formed. This is true because different
nuclear-synthesis processes occur in stars of different
masses, which evolve at different rates, and so eject
the heavy elements they produce into the interstellar
medium on different time scales. As a result, the
relative elemental abundances in the atmospheres of
subsequent generations of stars with various ages can
be used to trace the history of the enrichment of the
stellar--gas system in various chemical elements.

As typical representatives of the Galactic thin disk,
open clusters are often used to analyze the spatial
structure and chemical properties of the thin disk.
However, studies of the elements of their Galactic orbits
and their metallicities suggest that the population
of open clusters is not homogeneous, and includes
clusters formed as a result of interactions of high-velocity,
massive clouds with interstellar material in
the thin disk, and also clusters formed from interstellar material
that has fallen onto the disk from outer regions
of the Galaxy (see \cite{1}-\cite{4} and references therein).
This same conclusion was drawn in connection with
the detection of clusters with very elongated orbits extending
far from the Galactic plane having low metallicities
characteristic of older Galactic subsystems—
the thick disk and even the halo. The possibility of
forming open clusters with high, non-circular orbits
and low metallicities as a result of this mechanism for
initiating star formation has been demonstrated and
theoretically modeled (see, e.\,g., \cite{5}).

The chemical composition of high-velocity clouds
differs from that of the interstellar matter in the thin
disk, making it of interest to compare the the relative elemental 
abundances of clusters and field stars,
which could differ. It is also important to trace possible
differences in the chemical compositions of clusters
with high velocities or low metallicities, uncharacteristic
of thin-disk stars. Prompted by their interest
in the detailed chemical compositions of clusters
with different spatial and kinematic properties, Jacobson
et\,al. \cite{6} compared the relative abundances
of three $\alpha$-elements in 12~open clusters of the outer
disk, at distances from the Galactic plane $|z| > 1$ kpc,
and 11~clusters located closer to the plane, but found
no appreciable systematic differences. No evidence
was found in the chemical compositions of the cluster
stars in the outer disk that they were members of the
thick disk. Moreover, it was noted that two clusters
associated kinematically with the remnants of a
dwarf elliptical galaxy in Sagittarius (Sgr DSph) \cite{7}
likewise do not display appreciable differences in the
relative abundances of the studied elements for stars
of the field and the thick and thin Galactic disks.
Unfortunately, as is also true of other similar studies,
the sample of clusters in this study is too small to
enable statistically significant detections of systematic
differences, which are unlikely to exceed the uncertainties
in the elemental abundances in individual
clusters.

The aim of the current study is a comparative analysis
of the dependences of the relative abundances of
$\alpha$-elements on metallicity, age, Galactocentric distance,
and the parameters of the Galactic orbits, for
both open clusters and field stars of the thin disk,
and also for different groups of open-cluster populations.
In order to estimate the statistical reliability
of the results, we must first collect the most accurate
published spectroscopic measurements of the
abundances of elements produced in various nuclear-synthesis
processes in stars located in as large a
selection of open clusters as possible.


\section {INPUT DATA}

We used version 3.4 (May 2014) of the catalog
\cite{8}, which contains data for 2167 open clusters, for
918~of which proper motions and radial velocities are
available. This version of the catalog is more than
50\,\% larger than the version we used in our earlier
studies \cite{3, 4}. The Galactic coordinates, distances,
proper motions, and radial velocities were used to calculate
the Cartesian and cylinrical components of the
cluster velocities. We then used these data to calculate
the elements of the Galactic orbits of the clusters.
The motion of the Sun relative to the Local Standard
of Rest (LSR) was taken to be 
$(U, V, W)_{\odot}  = (11.1, 12.24, 7.25)$~km/s \cite{9}, 
the solar Galactocentric
distance to be 8.0~kpc, and the rotational speed of
the LSR to be 220~km/s. The orbits were calculated
for a standard model of the Galaxy having rotational
symmetry [10], consisting of a central region having
two spherical components with a total mass of
$1.9*10^{10} M_{\odot}$, a disk with mass $3.842*10^{10} M_{\odot}$ 
consisting of three Miyamoto-Nagai potentials and
three terms in the Miyamoto-Nagai potential imitating
a thin layer of gas in the disk, and a spherical halo
described by a logarithmic potential. The integration
was carried out over a time interval of $10^9$~years using
a fourth-order Runge-Kutta method that conserves
energy and angular momentum. This yielded the
apogalactic $R_a$ and perigalactic $R_p$ distances, the
maximum distance of points on the orbit from the
Galactic plane $Z_{max}$, and the orbital eccentricity 
$e = (R_{a} - R_{p})/(R_{a} + R_{p})$. These parameters were calculated
as their maximum values during the $10^9$~year
interval. (See \cite{8} for a discussion of the uncertainties
in the estimates of the ages, distances, velocities, and
Galactic orbital elements of the clusters.)

The main sources of metallicities were data from
the catalogs of Heiter et\,al. \cite{11} and Dias et\,al.
\cite{8} [version 3.4, May 2014]. Since Heiter et\,al. \cite{11}
rigorously selected the data used and averaged the
metallicities over several measurements obtained using
high-resolution spectra, their study yielded high-quality
data. It is primarily these metallicities that we
give in our table. However, for six of the clusters from
\cite{11}, new data became available after the publication
of this catalog. We obtained the metallicities of five
of these (Collinder~110, Collinder~261, NGC~2477,
NGC~2506, NGC~5822) by averaging the data of \cite{11}
and \cite{12}. For another cluster (NGC 1901), we calculated
the mean metallicity for the data of \cite{8, 11} (the
metallicity for this cluster in \cite{8} was determined in
2014). The metallicities in \cite{8} were collected from various
sources, and the [Fe/H] values were determined
using both spectroscopic and photometric methods.
We added all spectroscopic values from \cite{8} to our
sample, as well as photometric [Fe/H] measurements
obtained after 2010. Moreover, we took the meanweighted
photometric metallicities of \cite{13}, and then
added all values that fell into only one of the lists. The
photometric metallicity of Berkeley 60 determined before
2010 was obtained by averaging the data from
these two catalogs. We adopted spectroscopic metallicities
from \cite{14,16} and photometric determinations
from \cite{17} for four clusters. In all, we have collected
metallicities for 346~clusters. The mean uncertainty
in the metallicities is $\varepsilon{\rm [Fe/H]} \approx 0.10$.

We supplemented these data with the relative
abundances of 14~elements in stellar clusters taken
from 109~studies published from 1991 through 2015.
The elemental abundances in open clusters from
all the sources used were determined using high resolution
spectra, and were obtained mainly for
atmospheres of red giants. Our analysis of the spectra
assumed local thermodynamical equilibrium (LTE),
but deviations from LTE were taken into account in
some studies, primarily for oxygen and magnesium.
Most of the studies used Kurucz model stellar atmospheres.
Since our aim is to study the chemical
evolution of the Galactic thin disk, we restricted
our analysis to abundances of elements that are
expected to undergo virtually no changes associated
with nuclear processes in the envelopes of the studied
stars. Such elements include, in particular, the б
elements (O, Mg, Si, Ca, and Ti), the iron-peak
elements (Fe), and elements produced by slow (Y, Ba,
La, Ce, Nd, and Zr) and rapid (Eu) neutron capture;
see \cite{18} for a justification of this selection. To these
we added two other elements with odd numbers of
protons: Na and Al. The ratios [el/Fe] for the clusters
were found from the abundances [el/H] and [Fe/H]
contained in the original references, and not the mean
metallicities we have presented here.

The mean number of studied stars in a cluster is
seven, and the most probable number is four. The
abundances were determined using one star in 16
cases; in only 6~clusters was this the only determination,
while [el/Fe] values are available in other studies
in the remaining cases. The maximum number of
stars in a cluster analyzed in a single study is 76
(NGC~2243). The elemental abundances for 246
Hyades stars are presented in 15~studies, and for 105
stars in NGC~2682 in 13~studies. A large number of
studied stars in a cluster appreciably increases the accuracy
with which their mean elemental abundances
can be determined. The statistics for the elemental abundance
determinations in the open clusters and
their uncertainties are presented in Table~1.

We calculated the mean-weighted abundances for
clusters with two or more abundance determinations
for each of the studied elements. The weighting
coefficients were taken to be inversely proportional
to the stated uncertainties; as can be seen in 
Table~1, the corresponding mean values for all elements
are in the range $\varepsilon[el/Fe] = (0.06-0.09)$. However,
the stated uncertainties for some elements in several
studies can sometimes exceed 0.3~dex. These
were assigned lower weights, but were included in
the catalog unchanged if they were the only available
measurement. Fortunately, as was noted above,
there are few such uncertain measurements, and they
were excluded from consideration if they appreciably
influenced the dependences shown by the remaining
data. If uncertainties were not indicated, we assumed
uncertainties of 0.2 for data published before 2000 and
of 0.1 for data published after 2000.

To estimate the similarity of the abundances of
each element obtained in different studies, we analyzed
the distribution of the deviations of the published
relative abundances for a given cluster from
the calculated mean-weighted values; several dozen
overlapping values were found for each of the elements
studied, apart from cerium and neodymium.
All the resulting histograms are fit well by normal
distributions, indicating that the errors are random.
On average, the dispersions are $\sigma[el/Fe] = 0.10 \pm 0.01$, 
with barium and europium displaying the largest
dispersions. Thus, the external uncertainties in the
relative abundances in clusters are somewhat higher
than the values stated in the original papers, on average.

Our resulting catalog contains 346 open clusters
with known metallicities, for 272 of which total
spatial velocities have been calculated. Table~2
presents a fragment of the full table (catalog), which
is accessible in electronic form at the site ftp://
cdsarc.u-strasbg.fr/pub/cats/J/AZh, and contains
all the derived parameters for these clusters.
The columns of this table give (1) the name of the
cluster; (2)--(3) the Galactic coordinates ($l$, $b$); (4) the
heliocentric distance d; (5)--(7) the coordinates (x, y,
z) in a right-handed Cartesian coordinate system; (8)
the Galactocentric distance $R_{G}$; (9)--(11) the three
calculated spatial-velocity components $V_{R}$, $V_{\Theta}$, $V_{Z}$ 
in cylindrical coordinates, where $V_{R}$ is directed toward
the Galactic anti-center, $V_{\Theta}$ in the direction of the
Galactic rotation, and $V_{Z}$ toward the Galactic North
pole; (12) the eccentricity of the Galactic orbit $e$; (13)
maximum distance of points in the orbit from the
Galactic plane $Z_{max}$; (14)--(15) the apogalactic $R_a$
and perigalactic $R_p$ distances of the orbit; (16) the
cluster age; (17)--(18) [Fe/H] and references to the
sources for the metallicities; (19)--(32) the relative
abundances of 14~elements [el/Fe]; and (33) references
to the sources of the chemical compositions.
The coding for the [Fe/H] and [el/Fe] reference numbers
are presented in electronic form in the catalog.

For comparison, we considered uniform abundance
determinations for the same elements based
on high-resolution spectra of 212~field dwarfs and
171~field red giants, since it is such high-luminosity
stars that were usually observed in the clusters. All
these measurements were obtained by a single team
of authors \cite{19}--\cite{22}. We used Kurucz models to determine
the elemental abundances in these stars. The
uncertainties in the abundances of all elements do
not exceed 0.15~dex. We were able to find the distances,
proper motions, and radial velocities for all
the giants in the new Hipparcos catalog \cite{23}. We
used these data together with the Galaxy model of
Gardner and Flynn \cite{10} to calculate the elements of
the Galactic orbits. Since masses for the giants are
also available in the literature, we used these to calculate
the giant ages in years, using the relation for the
main-sequence lifetime for stars with solar metallicity
$lg t = 10 - 3.6 lg(M/M_{\odot}) + lg^{2} (M/M_{\odot})$ \cite{24}.
We also considered for comparison uniform estimates
of the abundances of the same elements based on
high-resolution spectra of 221~field Cepheids from
\cite{18}, located at the same distances as the open clusters.
Since the Cepheid elemental abundances were
determined by the same authors as those determining
the field dwarfs and giants, the uncertainties in
$\varepsilon{\rm[el/Fe]}$ do not exceed those for the abundances in
these other field stars. See \cite{18} for more detail on the
parameters of these comparison stars.

In our current study, we investigated the behavior
of the relative abundances of $\alpha$-elements using our
compiled catalog; the properties of other elemental
abundances will be considered in subsequent papers.

\section {METALLICITY DEPENDENCE OF THE RELATIVE ABUNDANCES
OF $\alpha$ ELEMENTS}. 

The variation in the relative abundances of some
elements [el/Fe] with increasing metallicity can be
interpreted in terms of the chemical evolution of a
stellar-gaseous system only if the overall abundance
of heavy elements in the system increases with time,
on average; i.\,e., if the metallicity is a statistical indicator
of age. Studies of the metallicity in the Galactic
thin disk show that stars of any age display large
scatters in their heavy-element contents, hindering
identification of a relation between age and metallicity
in a subsystem. Many studies have reported an absence
of such a relation (see, e.\,g., \cite{25}--\cite{28}). However,
this conclusion has been questioned in other studies,
which have suggested that a relation is present, but is
blurred by the uncertainties in the stellar ages and the
radial migration of stars (see, e.\,g., \cite{29}--\cite{33}).

A complete sample of 14 000 F--\,G\,dwarfs of the
thin disk located within 70~pc of the Sun was compiled
in \cite{34}, based on the catalog \cite{27}. Selection
effects, systematic effects associated with random
errors in the age determinations, and the effects of
radial migration of stars were analyzed. It was shown
in \cite{34} that, during the first several billion years of the
formation of the thin disk, the interstellar material incident
on the disk possessed approximately the same
heavy-element abundance ($\langle{\rm [Fe/H]}\rangle \approx -0.2$) and a
low degree of homogeneity, but the dispersion in the
metallicity $\sigma{\rm [Fe/H]}$ decreased smoothly from $\approx0.22$
to $\approx0.13$ with age. However, approximately four to
five billion years ago, the mean metallicity began to
systematically increase, while preserving the same
dispersion. A similar behavior for the age dependence
of the metallicity is also displayed by thin-disk stars
with uniform spectroscopic iron abundances from the
compiled catalog \cite{35}, which contains about 900 F--\,G\, 
dwarfs and ages estimated from the chromospheric
activity of the stars.

It is proposed in \cite{34} that the low star-formation
rate in the initial stages of formation of the Galactic
thin disk suddenly increased about four to five billion
years ago. This means that the metallicity in the thin
disk over the past $\sim 5$ billion years can be considered
to be related to age, so that the observed differences
in the relative abundances of ¦Б elements in the open
clusters compared to the field stars can be interpreted
in terms of the chemical evolution of the interstellar
matter. This makes it interesting to compare the
metallicity dependences of the relative abundances of
$\alpha$-elements for the open clusters and the field stars.

According to modern views, $\alpha$-elements are
primarily synthesized in the cores of nassive stars
($ M> 10M_{\odot}$) in late stages of their evolution, and
are ejected into the interstellar medium when they
explode as Type II supernovae (see, e.\,g., \cite{34,37}).
Type II supernovae also give rise to some iron-peak
elements, but iron-peak elements are produced
primarily in Type Ia supernovae, which correspond
to the final stage in the evolution of close binary
stars with masses $< 8M_{\odot}$ (see, e.\,g., \cite{38}). Since
the evolution times of massive stars do not exceed
$\approx 30$ million years and massive Type Ia supernovae
occur only after $\approx1-1.5$~ billion years, the relative
abundances [$\alpha$/Fe] of new generations of stars should
decrease as the metallicity increases. The lower the
star-forming rate in a stellar-gaseous system, the
lower the metallicity at which this decrease begins,
and the steeper it is.

Another important property of the abundances of
$\alpha$-elements can also be used to analyze the star 
formation history. According to current views, only
oxygen, and possibly magnesium, are produced entirely
in massive stars, while the two other $\alpha$-elements
we are using are often produced in less massive stars.
(Titanium is often not considered an $\alpha$-element, and
we do not consider it here.) Theoretical computations
show that the output of so-called primary $\alpha$ elements
(O, Mg), synthesized in hydrostatic processes in carbon
cores and explosively through Ne/C burning in
layer sources in pre-supernovae, increases with the
mass of the pre-supernova (see, e.\,g., \cite{39,40}). Therefore,
the observed ratios [$\alpha$/Fe] in the atmospheres of
new-generation stars will increase as the initial mass
function is shifted toward higher masses.

A more trustworthy indicator of the maximum
masses of supernovae is the ratio of the primary and
secondary (Si, Ca) $\alpha$-elements. However, the fact
that secondary $\alpha$-elements are also produced in lowermass
stars must be taken into account \cite{41}. These
properties of the $\alpha$-element abundances enable comparison
of the maximum masses of supernovae that
have enriched clouds producing clusters with various
origins, as well as the interstellar medium of the thin
disk.

Figure~1 shows the metallicity dependences of the
relative abundances of each of the four $\alpha$-elements
considered in open clusters (circles) and in field red
giants (x's). The relative oxygen abundances for the
field giants show a moderate increase with decreasing
metallicity, while the slope of the corresponding
relation for the open clusters is much lower, although
it differs from zero by more than $3\sigma$ (the correlation
coefficients are $r = -0.72 \pm 0.10$ with the probability
that the correlation arose by chance being $P_{N} \ll 1$\,\%,
and $r = -0.41 \pm 0.13$ with $P_{N} < 2$\,\%, respectively).
As a result, the relative oxygen abundances for most
of the clusters with metallicities higher than the solar
value are approximately the same as those for the field
giants; roughly half the lower-metallicity clusters lie
below the band occupied by field stars. Simultaneously,
a substantial scatter in the [O/Fe] values is
observed for the clusters, especially for those with
metallicities below the solar value.

A similar difference in the slopes is observed for
magnesium, but the relative Mg abundances for most
of the clusters and field giants coincide at small
metallicities. Some clusters have uncharacteristically
low or high [Mg/Fe] ratios for field giants. At the
same time, the overwhelming majority of clusters
with higher than solar metallicities lie above the mean
values for field giants (i.\,e., above the corresponding
fitted lines).

The dependences for silicon and calcium are very
similar for the two types of objects due to the appreciable
decrease in the slopes for these secondary $\alpha$-elements
for the field stars. For both elements, the scatter
of the points in the diagrams is appreciably higher
for the clusters than for the field giants. Given that the
field giants currently located in the solar vicinity were
born at greater Galactocentric distances, comparable
to the distances to the clusters (Fig.~4), and the range
of ages for the two types of objects are the same
(Fig.~3), this suggests that the smaller scatter in
the relative abundances of $\alpha$-elements displayed by
the field giants indicates that the interstellar medium
from which they formed was more uniform than the
medium from which the open-cluster stars formed.

Let us now consider another property of all these
diagrams. We showed in \cite{3,4} that open clusters can
be divided into two groups according to the orbital
parameter $(Z^{2}_{max}+4e^{2})^{1/2}$ proposed in \cite{42}, with the
division being at 0.35 (here, $e$ is the eccentricity of
the Galactic orbit and $Z_{max}$ the maximum distance
of the orbit from the Galactic plane in kiloparsec).
We referred to clusters with low orbital parameters as
``Galactic'' and to those with high orbital parameters
as ``peculiar''. The peculiar group also includes clusters
whose orbital elements are unknown, but that are
located more than 350~pc from the Galactic plane;
i.\,e., they satisfy the criterion above for peculiar clusters.
Note that the peculiar group also includes clusters
with low orbital parameters but uncharacteristically
low metallicities for thin-disk objects, ${\rm [Fe/H]} \approx < -0.2$. 
However, we did not find published elemental
abundances for these clusters. As a result, all the
clusters in the sample considered here with high,
elongated orbits (i.\,e., with high orbital parameters)
are peculiar, and all the clusters with low, circular
orbits are Galactic.

Following Vande Putte et\,al. \cite{2}, we proposed
that clusters with high, elongated orbits have an ``unusual''
origin. In \cite{3,4}, we used the orbital elements
calculated in \cite{1}, while here we have calculated the orbital
elements ourselves, but using another model for
the Galaxy; this became possible due to the increasing
number of clusters with measured radial velocities.
This led to a modest increase in the sizes of the
calculated orbits, so that the division of the clusters
into the two groups corresponds to the critical value
for the orbital parameter $(Z^{2}_{max}+4e^{2})^{1/2} = 0.40$ and
a distance from the Galactic plane $|z| = 400$~pc. In
all the diagrams presented here, the two groups of
clusters are denoted using different symbols: peculiar
clusters by grey circles and Galactic clusters by
circled dots. Our analysis has shown that none of
the dispersions of the relative abundances of the $\alpha$-
elements display significant differences between the
two groups. However, the relative abundances of
each of the $\alpha$ elements are higher, on average, for
the peculiar clusters than for the higher-metallicity
Galactic clusters, which is a consequence of the general
negative correlation between [el/Fe] and [Fe/H].
In order to reduce the random errors in the abundances
of each element, we averaged the abundances
of all four $\alpha$-elements (O, Mg, Si, and Ca). Since, as
follows from Fig.~1, the relative abundances of each
of the four $\alpha$-elements for each object display different
deviations from the solar values, when aiming
to reveal undistorted dependences, the data for the
clusters and field stars should be averaged over all
four $\alpha$-elements simultaneously. Our analysis shows
that the distribution of the deviations of the relative
abundances of each $\alpha$-element from the calculated
values of [$\alpha$/Fe], constructed using 53~clusters with
four determinations of the $\alpha$-elements, is described
well by a normal law with dispersion у
$\sigma[\alpha/Fe] = 0.09 \pm 0.01$, which we take 
to be an estimate of the
uncertainty in this quantity. Although this estimate
is comparable to the discrepancies for individual elements
indicated in Table~1, the reliability of the mean
elemental abundances is generally higher, since each
was produced mainly in a single process.

Figure~2 presents the metallicity dependences of
the averaged relative abundances of $\alpha$-elements for
the open clusters and field red giants and dwarfs. The
sequence of field dwarfs and giants, and the corresponding
fitted lines, essentially coincide (the fitted
line for the dwarfs is not shown, to avoid cluttering
the figure). A comparison of Figs.~1 and~2 shows that
averaging the four elements for the orbits appreciably
decreases the scatter in the relative abundances for a
fixed metallicity.

It follows from Fig.~2a that the correlations between
the relative abundances of the $\alpha$-elements have
high significances for both the clusters and field giants
($r = 0.59 \pm 0.06$ and $0.88 \pm 0.03$, with $P_{N} \ll 1$\,\% in 
both cases), with the slope of the dependence
for the clusters being significantly lower than that for
the field giants (see Fig.~2a), while the scatter of the
[$\alpha$/Fe] ratios are higher. This trend is due exclusively
to the behavior of the clusters with high, elongated
orbits, while the Galactic clusters essentially lie in the
band occupied by the field stars. However, the scatter
of the [$\alpha$/Fe] ratios within both groups of clusters are
roughly the same. The slope of the dependence for the
peculiar clusters is even smaller: $-0.13 \pm 0.04$ (this
fitted line is not shown in the figure). The dependences
for the clusters and field stars intersect near
the solar metallicity and the mean relative abundance
of $\alpha$-elements.

Thus, lower-metallicity clusters with high, elongated
orbits display lower [$\alpha$/Fe] ratios, while the corresponding
values in higher-metallicity clusters are
slightly higher than those for field stars. For example,
17~of 19~clusters with high, elongated orbits and
metallicities ${\rm [Fe/H]} < -0.1$ lie below the fitted line for
field stars (note that the abundance of $\alpha$-elements for
the cluster NGC~2266, which lies above this line, was
determined from data for only one star in one study,
and the other cluster lying above the line, Berkeley
21, has been suggested by some authors to have a
very low metallicity). In contrast, only~3 of~11 such
clusters with high metallicities are located below this
line (and those are within 0.03~dex of the line). It is
implausible that this came about due to a systematic
reduction/enhancement in the abundances of $\alpha$-elements
in low/high-metallicity clusters.

The mainly low ratios [$\alpha$/Fe] for low-metallicity
clusters come about purely due to the low abundances
of the primary $\alpha$-elements, as is clearly visible in
Fig.~2b, which presents the metallicity dependences
of the ratios of the abundances of primary and secondary
$\alpha$-elements for clusters and field stars. At
low metallicities, the [O, Mg/Si, Ca] ratios for most
clusters are appreciably lower than those for field
giants, while the opposite is true at high metallicities.
The Galactic clusters also display different ratios
of primary to secondary $\alpha$-elements from the field
stars, but not from metal-rich clusters with high,
elongated orbits. As a result, solar values for the
ratios of these elements are observed for clusters of
any metallicity; i.\,e., there is essentially no correlation
($r = -0.19 \pm 0.09$ with $P_{N} \approx 17$\,\%). At the same
time, the field giants show a clear, highly significant
anti-correlation ($r = -0.86 \pm 0.05$ with $P_{N} \ll 1$\,\%).
The well defined dependence of [O, Mg/Si, Ca] on
[Fe/H] for the field giants could come about due to
the fact that, with time, the interstellar medium of the
thin disk begins to become enriched with secondary 
$\alpha$-elements produced in Type II and Type Ia supernovae
\cite{41}, which simultaneously lead to an increase in its
metallicity.

Figures~2c and~3d show [$\alpha$/Fe] versus [Fe/H] for
the primary and secondary $\alpha$-elements, respectively.
Both correlations for the clusters are highly significant
($r = -0.51 \pm 0.09$ and $-0.41 \pm 0.08$, respectively,
with $P_{N} \ll 1$\,\% in both cases). The slopes
of the fitted lines for the primary and secondary $\alpha$
elements for the clusters are nearly the same, while
those for the field stars differ appreciably. As a result,
the sequence of secondary $\alpha$-elements for the clusters
and field giants essentially coincide in Fig.~2d. It
is striking that, in both diagrams, clusters with low,
circular orbits lie mainly in the bands occupied by field
stars. As a result, the slopes of the two fitted lines for
clusters with high, elongated orbits in Figs.~2c and~3d
are still lower. Thus, the differences between clusters
with high, elongated orbits and field giants become
still larger for primary $\alpha$-elements, and smaller for
secondary $\alpha$-elements. Clusters that deviate substantially
from the sequence for the field stars are
shown in a different font in Fig.~2c. The bold font
denotes clusters in which the oxygen and magnesium
abundances were determined in several studies, and
regular font those in which these abundances were
determined in only one study, but using several stars
(only the abundances in Be~31 were determined using
only one star).

When the deviant clusters delineated using regular
font are excluded, the fitted slope for the remaining
clusters remains the same, within the uncertainties.
Thus, the differences in the slopes of the metallicity
dependences of the relative abundances of primary
$\alpha$-elements for clusters with high, elongated orbits
and field giants are unlikely to be due to systematic
errors in the oxygen and magnesium abundances of
the cluster stars. (Note that the procedure used to test
the stability of the fitted lines obtained by excluding
deviant points was carried out for all the diagrams in
Figs.~1 and~2, and the fitted slopes for the clusters
remained the same within the errors.)

\section {AGE DEPENDENCES OF THE METALLICITY AND RELATIVE
ABUNDANCES OF $\alpha$ ELEMENTS}.

{\bf Metallicity}. Figure~3a presents age--metallicity
diagrams for the open clusters and field red giants.
Since it was only possible to calculate a lower limit
for the ages of the red giants based on their masses,
which are presented in the original references with
accuracy to within one figure after the decimal place,
these data yielded discrete diagrams. This step was
equal to several billion years for giants with high
ages, i.\,e., low masses, while it decreased to tens
of millions of years for younger giants. In all, we
obtained nine distinct age values. As can be seen in
Fig.~3 and Fig.~4, all the age dependences considered
deviate appreciably from linear trends. Therefore,
when comparing the behavior of the age dependences
of the metallicities for open clusters and field giants, it
was most convenient to approximate the latter using
line segments joining the mean metallicities at the
nine age points, while the cluster dependences were
approximated using trends obtained from a running
average. The mean absolute uncertainties for the
cluster ages are presented in the figure, but the uncertainties
in the ages are reflected more accurately
by their relative uncertainties, which are $\approx30\,\%$, on
average \cite{3}.

Figure~3a shows that the behavior of the field red
giants is in good agreement with the behavior of the
field dwarfs in \cite{34} described above-the systematic
decrease in the metallicities before five billion years
is obvious, and the dependence becomes flat with
further increase in the age. (The one point near 1.7
bilion years that disrupts the smooth behavior of the
age dependence of the metallicity for the field giants
has a small mean-square error, but represents an
average over only five stars, and is unlikely to be
reliable.) In addition, a large scatter in the metallicities
exceeding the random uncertainties in [Fe/H] is
observed for both the field giants and dwarfs. Note
that the coincidence of the general appearance of the
age-metallicity relations for the field red giants and
dwarfs suggests that systematic effects associated
with uncertainties in the models for the red-giant
atmospheres have not led to appreciable distortion of
their derived ages. It is also significant that the dwarf
ages determined in \cite{34} were obtained both using
theoretical isochrones, which are likewise dependent
on theoretical atmospheres, and the stars' chromospheric
activity.

There is no correlation between age and metallicity
for the clusters as a whole ($r = -0.1 \pm 0.3$ with 
$P_{N} \approx 28$\,\%), although the mean metallicity varies in a 
complex fashion with increasing age. In contrast to the
age-metallicity dependence for the field stars, which
reflects evolutionary variations in the chemical composition
of the interstellar medium in the thin disk,
this dependence for the clusters is most likely due to
differences in the lifetimes of clusters with different
orbits \cite{4}. As the figure shows, the overwhelming
majority of high-metallicity clusters (${\rm [Fe/H]} > -0.2$)
have low, circular orbits (denoted by circled points
in all diagrams); due to their small residual velocities,
they remain near the spiral density wave that
stimulated their birth over long times, leading to the
rapid disruption of such clusters. The bulk of clusters
with such orbits ($\approx90$\,\%) have ages of less than 0.5
billion years. In contrast, almost all the clusters
with high ages for which the total velocities have
been measured have high, elongated orbits (shown
by light filled circles in the diagrams). It is believed
that clusters with high, elongated orbits and high
metallicities were formed from interstellar material of
the thin disk that received a high initial velocity when
it was born due to globular clusters intersecting the
Galactic plane, while lower-metallicity clusters with
similar orbits obtained both their momenta and their
low metallicities from giant molecular clouds \cite{4} (see
also references therein). As a result, clusters that end
up in high, elongated orbits spend a large fraction of
their time far from spiral arms, and are not subject to
further disruption.

This suggests that, when total-velocity measurements
become available for all clusters, old clusters
will probably end up in high, elongated orbits.
Figure~3a shows that a rapid increase in the mean
metallicity with increasing age is observed for clusters
with ages $<0.3$~billion years. (The behavior of clusters
with small ages can be seen in more detail in Fig.~1b of
\cite{4}, which presents a similar age-metallicity diagram
for clusters using a logarithmic age scale.) This is
associated with the growth in the fraction of low-metallicity
clusters among the youngest clusters. In
particular, only one-quarter of clusters younger than
20~million years with known metallicities have metallicities
higher than the solar value. At 0.3~billion
years, the mean metallicities of the clusters and field
giants agree within the uncertainties; further, the
mean metallicities of clusters begin to decrease much
more rapidly than those of field stars; at two billion
years, they have become roughly the same as those of
field giants older than $\approx4.5$~billion years.

The decrease in the mean metallicity of the clusters
is associated with a decrease in the fraction
of long-lived Galactic high-metallicity clusters with
increasing age. The metallicities of clusters older
than $\approx 2$ billion years display a very high dispersion
due to the absence of a concentration toward some
distinct metallicity value, since most high-metallicity
clusters whose orbits are located fully in the plane of
the Galaxy have already been disrupted. As a result,
long-lived clusters with ages of $0.5 - 4$~billion years
have lower metallicities, on average, than do field
stars. Among even older clusters, we again note a
tendency for an increase in the mean metallicity due
to the dominance of clusters with metallicities that
are uncharacteristically high for field stars of the same
age.

{\bf $\alpha$-elements}. Figure~3b plots the age versus the
relative oxygen abundances for field red giants and
open clusters. The mean [O/Fe] ratios of the field
giants initially increase, but then become constant
within the uncertainties after $\approx 4.5$~billion years. This
ratio also initially increases for the clusters, but only
to ages of roughly two billion years, after which
there is even some decrease with age. The sequence
of clusters everywhere lies below the points for the
field giants, although the two are very close at ages
$<2$~billion years. Here also, the slope $s$ of the age
dependence of [O/Fe] for the clusters is essentially
zero ($s_{clust} = 0.004 \pm 0.009$ billion years~$^{-1}$), while
this slope is significantly different from zero for the
giants ($s_{giant} = 0.013 \pm 0.004$~billion years~$^{-1}$).

The age dependence of the relative magnesium
abundances in Fig.~3c behave somewhat differently.
As in the case with oxygen, the [Mg/Fe] ratios for
the field giants increase monotonically with age at
small ages, but cease to grow after $\approx 4.5$ billion years,
while the rapid growth continues only to $\approx 0.5$ billion
years in the clusters. In contrast to oxygen, the
two lines essentially coincide at large ages; before
$\approx 3$ billion years, the mean line for the clusters lies
above the analogous line for the field giants, with the
difference being outside the uncertainties. The
overall slopes for the magnesium dependence for the
field giants is significantly greater than zero 
($s_{giant} = 0.025 \pm 0.003$~billion years~$^{-1}$), 
while the corresponding slope for the clusters is zero 
within $3\sigma$ ($s_{oc} = 0.022 \pm 0.008$ billion 
years~$^{-1}$). The age dependences
of the relative abundances of silicon and calcium for
the clusters and field giants shown in Figs.~3d and
3e essentially coincide at small ages; further, the sequences
for the clusters lies outside the uncertainties
in the mean values below. The overall age variations
for both objects reproduce the behavior of the other
two elements, and both slopes for the clusters are
equal to zero within the uncertainties.

Without considering specific origins for the differences
in the behaviors of the different $\alpha$-elements,
which could be partially associated with both systematic
errors and real differences in their abundances,
let us consider the age dependence of the relative
abundances of the $\alpha$-elements averaged over all four
elements. Here also, the fitted slope for the field giants
is greater than zero, while the slope for the clusters
is zero within the uncertainties 
($s_{giant} = 0.014 \pm 0.002$~billion years~$^{-1}$ and 
$s_{oc} = 0.010 \pm 0.004$~billion
years $^{-1}$). Figure 3f shows that the dependences
for both types of object pass through the coordinate
origin; i.\,e., the current mean relative abundances of
the $\alpha$-elements are solar. These relations then grow
in roughly the same way until $\approx2$~billion years, but
the growth in the clusters stops at $[\alpha/Fe] \approx +0.07$,
while it continues in the giants to $\approx4.5$~billion years,
when $[\alpha/Fe] \approx +0.13$. The initial increase in [$\alpha$/Fe]
for the clusters was appreciably slower than for the
field giants, as was the case for the increase in the
metallicity in Fig.~3a.

Recall that, among the clusters with ages of more
than 2~billion years, only two (Collinder~106 and
Collinder~261) have low, circular orbits, while all the
remaining old clusters have elongated, high orbits.
These clusters probably formed from a mixture of
material that had undergone chemical evolution that
was different from the interstellar matter of the thin
disk, and [$\alpha$/Fe] values that differ from those for field
stars are often observed. As can be seen in Fig.~3f
and Fig.~4, the decrease in the mean [$\alpha$/Fe] ratios
with decreasing age arises purely due to the low relative
abundances of $\alpha$-elements in clusters with low,
circular orbits. The mean $\langle[\alpha/Fe]\rangle$ ratios in clusters
with elongated, high orbits do not depend on age
($s_{oc} = 0.005 \pm 0.006$~billion years~$^{-1}$).

\section {RELATIONSHIP BETWEEN THE RELATIVE ABUNDANCES
OF $\alpha$ ELEMENTS AND LOCATION IN SPACE} 

{\bf Radial gradients}. The abundances of most
chemical elements in the Galactic thin disk decrease
with Galactocentric distance. These radial gradients
can be traced most reliably using Cepheids, which
are bright and visible from large distances. The
elements of the Galactic orbits of nearby stars can
also be used to determine these gradients. In this
case, the dependence of the elemental abundances
on the maximum or mean orbital radii for the stars
can be studied (see, e.\,g., \cite{43}). The radial metallicity
gradient among field stars of the thin disk arises
due to the increase in the star-formation rate with
decreasing distance from the Galactic center \cite{44}.
The radial gradients can also be studied using data
for open clusters \cite{22, 45}. However, as is pointed out
in \cite{2, 4}, these include clusters that formed as a result
of interactions between the interstellar medium of the
thin disk and various massive objects -- high-velocity
clouds, dwarf galaxies, and globular clusters. This
leads to anomalous chemical compositions in these
clusters, compared to field stars of the thin disk, as
well as high, elongated orbits; therefore, their radial
elemental abundance gradients do not reflect the starformation
history in the thin disk.

Figure~4a presents the dependence of the metallicities
of open clusters on Galactocentric distance $R_{G}$
(six clusters that are located at the largest Galactocentric
distances fall outside this diagram). The
corresponding dependences for Cepheids and field
red giants of the thin disk from \cite{18} and \cite{20}-\cite{22}
respectively, are shown for comparison. Due to the
proximity of the giants, the maximum distances of
their orbits from the Galactic center ($R_a$) were used
for their Galactocentric distances. The fitted linear
slopes for all of these objects are approximately
the same, although the slopes for the clusters are
somewhat lower than those for the field stars, even
using their true positions (although this difference
formally lies within the uncertainties). However, if
we use the apogalactic radii of the orbits, the slopes
for the clusters become lower, and the radial gradient
for the clusters becomes somewhat smaller than the
gradient for the field giants 
($d{\rm [Fe/H]}/dR_{a} = -0.01  \pm 0.01$~kpc$^{-1}$
versus $0.05 \pm 0.01$~kpc$^{-1}$). A smoothed
trend of the radius dependence of the metallicity for
the clusters obtained using a running average shows
that all the slopes are associated with a jump in the
metallicity near 9.5~kpc, reflecting the existence of
two populations of clusters with different properties.

Analysis of the elements of the Galactic orbits
presented in \cite{4} showed that the vast majority of
all clusters formed within a Galactocentric radius
of $\approx10.5$~kpc, and within $\approx 180$~pc of the Galactic
plane. However, due to their high initial velocities,
with time, peculiar clusters came to occupy a volume
comparable to that occupied by the oldest Galactic
subsystems. As a result, peculiar clusters, most having
uncharacteristically low-metallicities for field stars
of the Galactic disk, have ended up at large Galactocentric
distances. Since they formed from interstellar
material of the thin disk mixed with various fractions
of matter that had undergone different chemical evolution,
they could have different abundances of other
elements as well.

Although all $\alpha$ elements are ejected into the interstellar
medium mainly
by the same types of stars, their
relative abundances differ somewhat in subsequent
generations of stars. Let us consider the behavior
of the radial gradients derived for the averaged 
$\alpha$-elements. Figure~4b present the dependences of the
relative abundances of $\alpha$-elements in clusters, young
Cepheids, and field giants on their Galactocentric distances
(five clusters with large $R_G$ values fall outside
the diagram); $R_a$ was used here for the field giants.
Due to the insufficient number of clusters with known
chemical compositions, it is not possible to statistically
significantly detect the jump-like variation in
the relative abundances at $ \approx 9.5$~kpc seen in Fig.~4a,
although the deficit of clusters in the fourth quadrant
of the diagram is consistent with the existence of such
a jump.

Figure~4a shows that, as for the field stars, the
relative abundances of $\alpha$-elements for the clusters increase
with increasing Galactocentric distance. The
fitted slopes for the radial gradient of $\alpha$-elements are
lower for clusters than for field stars (although they
are formally equal within the uncertainties). However,
when the apogalactic orbital radii are considered,
the difference in the slopes for the clusters and field
stars become significant. The gradient $ d{\rm[\alpha/Fe]}/dR_G$
derived using the true positions of the clusters is
also smaller than the analogous gradient for field
Cepheids ($d{\rm [\alpha/Fe]}/dR_{G} = 0.011 \pm 0.003$~kpc$^{-1}$ and
$0.018 \pm 0.004$~kpc$^{-1}$, respectively). An increase in
[$\alpha$/Fe] with distance from the Galactic center is easy
to understand for the field stars. $\alpha$-elements are produced
in the cores of massive stars, which eject them
into interstellar space together with some amount
of iron when they explode as Type II supernovae.
After about a billion years, this matter begins to be
enriched in iron-peak elements, which are ejected by
less massive Type Ia supernovae, whose progenitors
have evolved by this time. Since the star-forming
efficiency increases with decreasing distance from the
Galactic center, on average, there are fewer metals
in more distant stars, leading to higher [$\alpha$/Fe] ratios.
The Galactocentric positions of the clusters are not
determined by the place of their birth, as is the case
with field stars, but instead by the velocity they receive
when they are born, while the chemical composition is
determined by the fraction of low-metallicity matter
acquired from the parent high-velocity cloud. As a
result, as a rule, clusters that are further from the
center have fewer metals.

{\bf Vertical gradients}. Figures~4c and~4d present
the dependences of the metallicity and relative abundances
of $\alpha$-elements in clusters on distance from
the Galactic plane ($|z|$), and in field giants on the
maximum distance of the orbits from the Galactic
plane ($Z_{max}$). As in the plot of $R_{G}$ versus [Fe/H],
the smoothed trend in the $|z|$ -- [Fe/H] diagram for
clusters displays a sharp drop in the mean metallicity
at distances 100 pc $< |z| <$ 200 pc, supporting the
existence of two populations of clusters. This diagram
also shows that the dependences of the metallicity
and relative abundances of $\alpha$-elements on distance
from the Galactic plane are determined by clusters
with high, elongated orbits, as was the case for the
dependence on the Galactocentric distances of the
clusters. The difference in the slopes of the dependences
for the clusters and the field stars is very significant
($0.06 \pm 0.02$ and $0.18 \pm 0.04$, respectively, for
Fig.~4d). If we use $Z_{max}$ for the clusters, as we did for
the field giants, the vertical gradients in the metallicity
and relative abundances of $\alpha$-elements for the clusters
become still lower (but remain significantly greater
than zero). Note that the clusters and field giants
behave very similarly in $e$ -- [Fe/H] and $e$ -- [$\alpha$/Fe]
diagrams ($e$ is the eccentricity) [3, Fig.~2b]). This is
precisely why it is sometimes convenient to consider
both of these characteristics of the spatial orbits using
a single orbital parameter, $(Z^{2}_{max}+4e^{2})^{1/2}$.

\section {DISCUSSION} 

Thus, we see that the primary $\alpha$-elements in open
clusters and field stars display different dependences
on metallicity, age, Galactocentric distance, and distance
from the Galactic plane. Since no origin for
the systematic distortions in certain elemental abundances
in cluster stars has been found, the detected
differences between the behavior of the mean relative
abundances of primary $\alpha$-elements in clusters and
field stars very likely testify to different chemical evolution
histories for the interstellar matter from
which these objects formed.

The vast majority of thin-disk field stars are genetically
related, since they were formed from the interstellar
matter of this Galactic subsystem; all dependences
displayed by the relative abundances of the elements
they contain reflect the chemical evolution of
this subsystem. On the other hand, according to [4],
only high-metallicity (${\rm [Fe/H]} >\approx -0.2$) clusters with
low, circular orbits ($(Z^{2}_{max}+4e^{2})^{1/2} < 0.40$), formed
from the same material as did the field stars. This
possibility is supported by the fact that these clusters
lie mainly within the band occupied by field stars of
the thin disk in all the diagrams.

However, the clusters display appreciably higher
metallicities and ratios of primary to secondary 
$\alpha$-elements (similar to the solar values) than do the
field stars, on average. This means that the chemical
compositions of the clusters and field stars are
somewhat different. Young clusters with low, circular
orbits, but with metallicities uncharacteristially low
for thin-disk field stars, could plausibly form, for example,
from matter from the Magellanic stream that
is partially mixed with the interstellar matter in the
thin disk. Unfortunately, the red-giant branch has
not yet formed in these very distant clusters, and it is
not possible to carry out a detailed chemical analysis
of the atmospheres of their dwarfs due to their low
brightness. Most of these clusters belong to the
Perseus stellar complex, as can be seen in Fig.~3c of
\cite{45}.

Clusters whose orbits are very elongated and extend
far from the Galactic plane almost always display
different dependences than do field stars. For
example, at less than solar metallicities, as a rule,
the primary $\alpha$ elements in clusters display lower [O,
Mg/Fe] ratios than do field giants and dwarfs of the
thin disk (Fig.~2c). All such low-metallicity clusters
with measured total velocities are in high, elongated
orbits. However, we can see that clusters with high,
elongated orbits at more high metallicity have 
no much lower  [O, Mg/Fe] ratios,
on average, while field stars display an appreciable
decrease in these ratios with increasing [Fe/H]. As a
result, the relative abundances of primary $\alpha$-elements
in most of these clusters are higher than those in field
giants of the same metallicity.

Note that the metallicity distribution for clusters
with high, elongated orbits is bimodal, with two
groups separated by a dip near ${\rm [Fe/H]} \approx -0.1$
(Fig.~5a). The parameters of these Gaussians determined
from a maximum-likelihood fit show that the
probability of erroneously rejecting alternative fits in
favor of a sum of two Gaussian curves is $< 5$\,\% (see the
method in \cite{46}). This bimodality provides appreciable
support for the inhomogeneity of this group of clusters.
Both metal-poor and metal-rich clusters with
high orbital indices are mainly older than 1~billion
years (see Fig.~3). The age-metallicity diagram in
Fig.~5b shows that both metal-rich and metal-poor
clusters with high, elongated orbits display weak
($P_{N} > 5$\,\% for both) correlations with opposite signs.
(Clusters with very low metallicities differing from the
mean value for clusters with high, elongated orbits
by more than $3\sigma$ were excluded when obtaining the
regression fits.)

Further, it follows from Figs.~4a and~4b that the
vast majority of low-metallicity clusters with high,
elongated orbits are at distances of $R_{G} > 9.5$~kpc,
while all metal-rich clusters are located at smaller
Galactocentric distances. Although the slopes and
positions of the [$\alpha$/Fe] vs. $R_G$ dependences for the
clusters and field giants are fairly close, their different
[$\alpha$/Fe] vs. [Fe/H] and [Fe/H] vs. $R_G$ dependences
suggest that this is a chance agreement. (Recall
that we are supposing that the current Galactocentric
positions of peculiar clusters are not associated with
their birth places, as is the case for field stars, but
rather with their initial velocities, which depend on
the velocities of the high-velocity clouds from which
they formed.) Finally, clusters with 
$(Z^{2}_{max}+4e^{2})^{1/2} > 0.40$ display appreciable 
correlations between the distance
from the Galactic plane ($|z|$), on the one hand,
and the metallicity and relative abundance of $\alpha$-elements,
on the other—the more non-circular the orbit
and the higher the orbit above the Galactic plane, the
lower the metallicity and the higher the [$\alpha$/Fe] ratio
(Figs.~4c,~4d). It is possible that, when such clusters
were born, they acquired an admixture of matter with
some other chemical composition, proportional to the
impulse obtained.

In our view, all these facts support the conclusion
that low-metallicity clusters with high, elongated
orbits contain an admixture of lower-metallicity
material with lower relative abundances of primary
$\alpha$ elements than the typical values for field stars; i.\,e.,
they formed during collisions between high-velocity
clouds and the interstellar medium of the thin disk. It
appears that the velocity dependence of the metallicity
is characteristic for the high-velocity clouds—
lower-metallicity clouds travel to the Galactic disk
from larger distances, and therefore intersect the
Galactic plane with higher speeds. Low [O, Mg/Fe]
ratios could come about in high-velocity clouds in the
presence of such low metallicities if the star-forming
rate in the star–gas system that gave rise to these
clouds is lower than the rate in the thin disk. In this
case, Type Ia supernovae, which had already begun
to explode in the interstellar medium, enriched the
material in iron, thereby reducing the [$\alpha$/Fe] ratio,
although the metallicity itself remained appreciably
below the solar value. Moreover, [O, Mg/Fe] ratios
in clusters that are lower than in field stars could
come about if the mass of Type II supernovae that
have enriched the interstellar medium from which the
high-velocity clouds were born was lower than in the
thin disk.

The appearance of high-metallicity clusters in
high, elongated orbits is usually associated with
collisions between globular clusters or the cores of
decayed dwarf galaxies and the interstellar matter
in the thin disk \cite{2} (see also references therein).
However, their higher relative abundances of primary
$\alpha$ elements compared to field stars of the same
metallicity suggests that the birth of these clusters
was stimulated by other high-velocity clouds with
high abundances of heavy elements and intermediate
speeds, which it is believed could form from returning
gas in Galactic ``fountains'' (see, e.\,g., [47, 48]).

It is supposed that clouds with intermediate velocities
are composed of internal Galactic gas, while
high-velocity clouds are composed of primordial gas.
Observations have indicated enhanced abundances of
oxygen (an $\alpha$-element) in such clouds, which can be
explained if they form as a result of Type II supernovae.
The possibility that these clouds are enriched
in material from collapsing stars is also suggested by
the fact that enhanced ratios of primary to secondary
$\alpha$-elements, compared to field stars, are observed
in clusters formed from such clouds. (These ratios
are probably substantially lower in field stars due to
additional enrichment of the interstellar medium of
the thin disk in secondary $\alpha$-elements from Type Ia
supernovae \cite{41}.)

Note that all the tendencies described here are
purely statistical, and clusters that deviate from the
generally observed behavior require individual study.

\section {CONCLUSION}

Thus, the distinguishing characteristics of the behavior
of the relative abundances of $\alpha$-elements in
open clusters support the conclusion that some of
the clusters formed during interactions between low metallicity,
high-velocity clouds and the interstellar
matter of the Galactic thin disk, as was proposed earlier
based on analyses of the elements of their Galactic
orbits and metallicities. The low ratios of primary
$\alpha$-elements in low-metallicity (${\rm [Fe/H]} < -0.1$)
open clusters with high, elongated orbits suggest that
these high-velocity clouds were most likely formed
from interstellar material in which either the starforming
rate was lower or the mass of Type II supernovae
was higher than near the Galactic plane.
On the other hand, we suggest that the formation
of high-metallicity clusters with somewhat less high
and elongated orbits is associated with clouds with
intermediate velocities that formed from internal gas
returning to the disk in Galactic fountains, and that
their enhanced relative abundances of primary $\alpha$-elements,
compared to field stars, are due to the fact that
an appreciable matter in these clouds is from massive
Type II supernovae.

Metal-rich (${\rm [Fe/H]} > -0.2$) clusters with low, circular
orbits form a separate group. These are mainly
appreciably younger than clusters with high, elongated
orbits, and are located at Galactocentric distances
comparable to or less than the solar value.
Their relative abundances of $\alpha$-elements are the same
as those in field giants. These properties support the
idea that they are fairly closely and genetically related
to matter of the Galactic thin disk. The least metal-rich
clusters with circular orbits are young enough
that their red-giant branches have not yet formed, and
they do not contain stars that can be used for spectroscopic
determinations of their elemental abundances.
Therefore, it is not possible to elucidate their nature
based on their chemical compositions. However, their
metallicities, which are uncharacteristically low for
Galactic disk stars, suggest the presence of matter
from the Magellanic stream in the interstellar matter
from which they formed.

Our results and conclusions are valid in the absence
of systematic shifts in the elemental abundance
determinations, and if our current understanding of
the sources of these elements is correct. Our subsequent
paper \cite{49} continues our study of the population
of open clusters, and presents a statistical analysis of
the relative abundances of neutron-capture elements
in clusters of various groups and in field stars of the
Galactic thin disk.

\section*{ACKNOWLEDGMENTS}

We thank N.O. Budanova for help in collecting
material on the chemical composition of open-cluster
stars. This work was supported by the Ministry
of Education of the Russian Federation (State contracts
No.3.961.2014/K and No.213.01-11/2014-
5, project code 26.63), and the Southern Federal
University (grant 213.01-2014/013-VG).

\renewcommand{\refname}{Список литературы}

\newpage

\begin{figure*}
\centering
\includegraphics[angle=0,width=0.99\textwidth,clip]{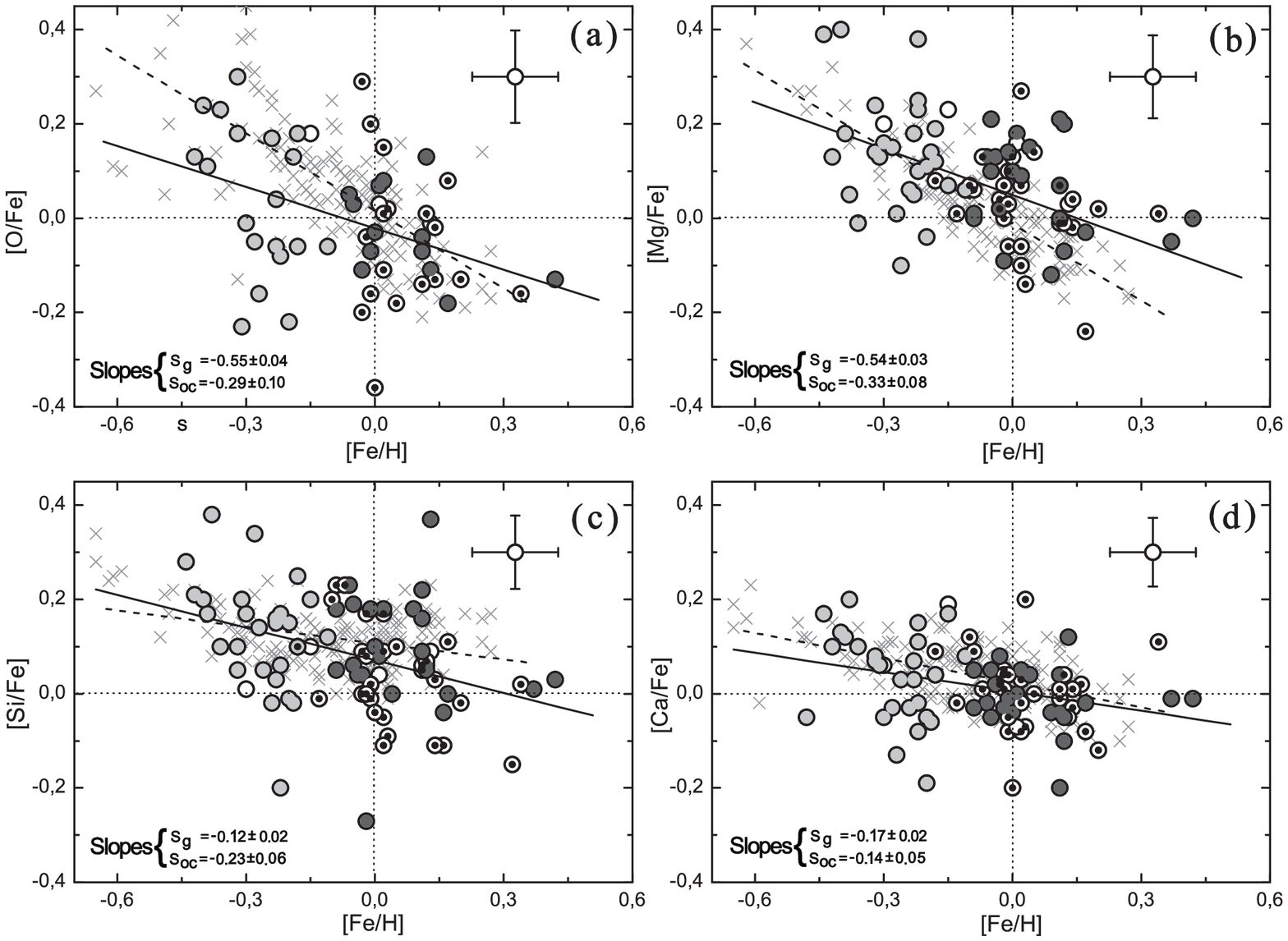}
\caption{Metallicity dependences of the relative abundances of 
         four $\alpha$ elements in open clusters (circles) and field 
         red giants (x's). The circled dots represent clusters with 
         low, circular 
         orbits (orbital indices $(Z^{2}_{max} + 4e^{2})^{1/2} < 0.40$ 
         and $|z| < 400$~pc).
         The light and dark gray filled circles represent 
         low-metallicity (${\rm [Fe/H]} < -0.1$) and high-metallicity clusters,
         respectively,with high, elongated orbits. The hollow circles 
         represent unclassified clusters. The solid lines show the 
         linear regression fits for the clusters, and the dashed lines 
         the corresponding fits for the giants. The slopes of the 
         fitted lines and their uncertainties are indicated. The circle 
         with bars shows the characteristic uncertainties for all the 
         elements. The dotted horizontal and vertical lines pass 
         through the solar values for the metallicities and [el/Fe].}
\label{fig1}
\end{figure*}

\newpage

\begin{figure*}
\centering
\includegraphics[angle=0,width=0.99\textwidth,clip]{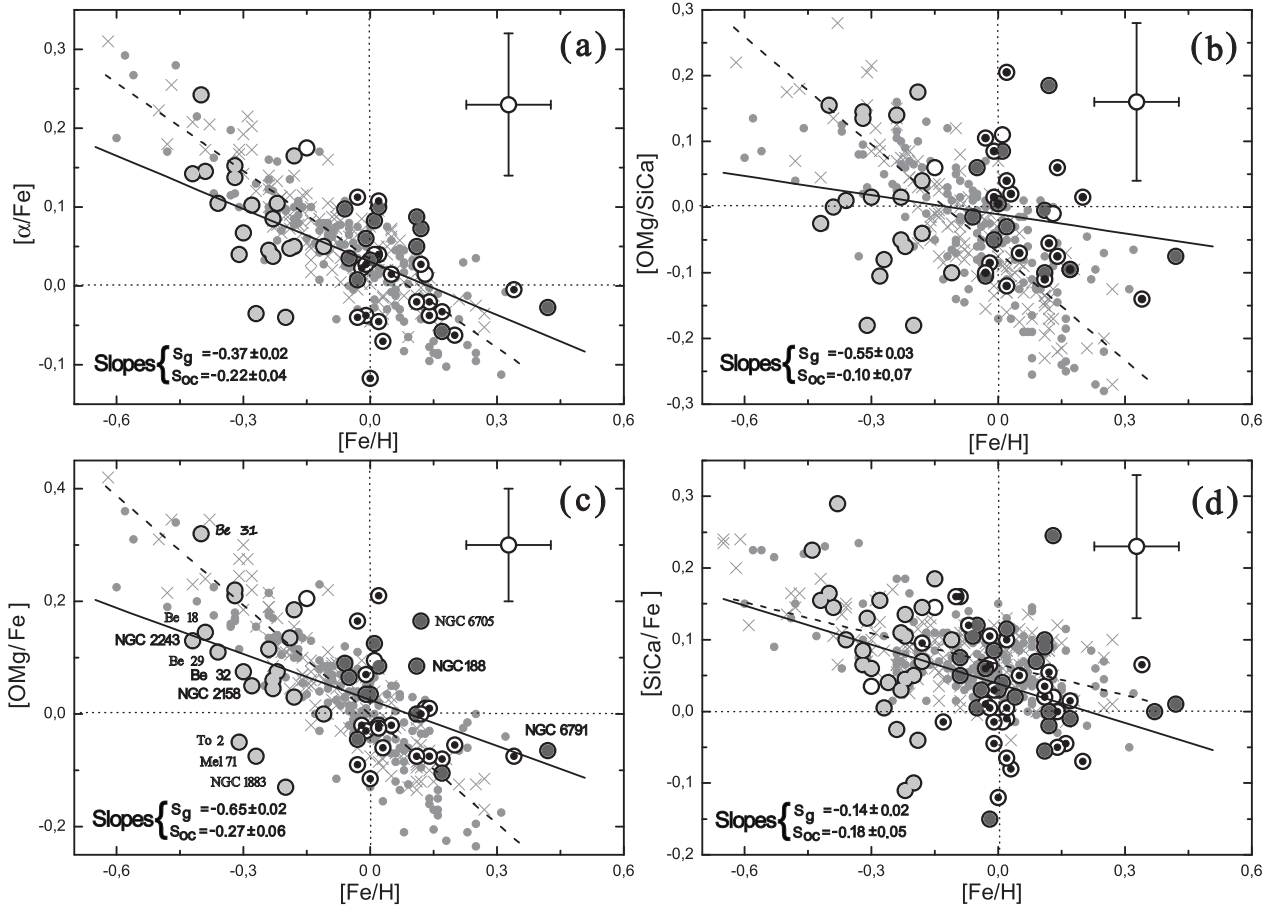}
\caption{Dependences of the (a) mean relative abundances of 
         $\alpha$ elements, (b) the ratio of the abundances of 
         the primary and secondary $\alpha$-elements, and the 
         relative abundances of the (c) primary and (d) secondary 
         $\alpha$-elements on metallicity, for open clusters 
         (circles), field red giants (x's), and field dwarfs 
         (points). The notation is the same as in Fig.~1. Clusters 
         with high, elongated orbits that have large deviations from 
         the plane occupied by the field giants are indicated in 
         panel (c); the abundances for those in bold font have been 
         determined in several studies, for those in normal font in 
         a single study using several stars, and for those in italic 
         font using data for a single star.}
\label{fig2}
\end{figure*}

\newpage

\begin{figure*}
\centering
\includegraphics[angle=0,width=0.95\textwidth,clip]{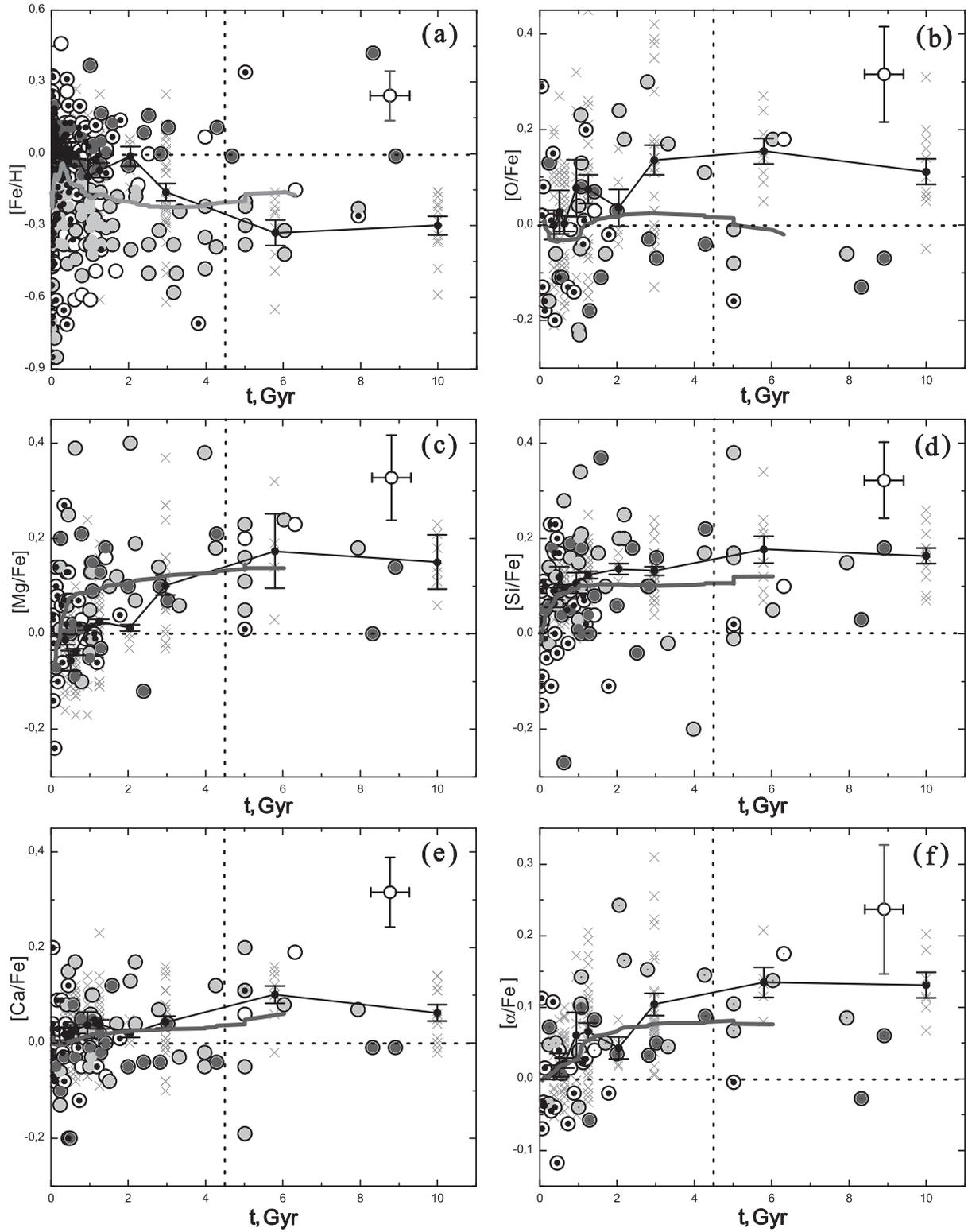}
\caption{Age dependences of the (a) metallicity, (b)-–(e) relative 
         abundances of four $\alpha$-elements, and (f) mean relative abundances
         of the $\alpha$-elements for open clusters (circles) and red giants
         of the field (x's). The smooth curves show fitted smoothed 
         trends obtained from a running average of the age dependences 
         of the clusters; the segmented curves show the mean 
         metallicities and $\langle{\rm[\alpha/Fe]}\rangle$ ratios 
         in nine narrow age bands 
         for field giants joined by line segments; the bars denote 
         the uncertainty in the mean values for the giants. The 
         remaining notation is as in Fig.~1.}
\label{fig3}
\end{figure*}

\newpage

\begin{figure*}
\centering
\includegraphics[angle=0,width=0.99\textwidth,clip]{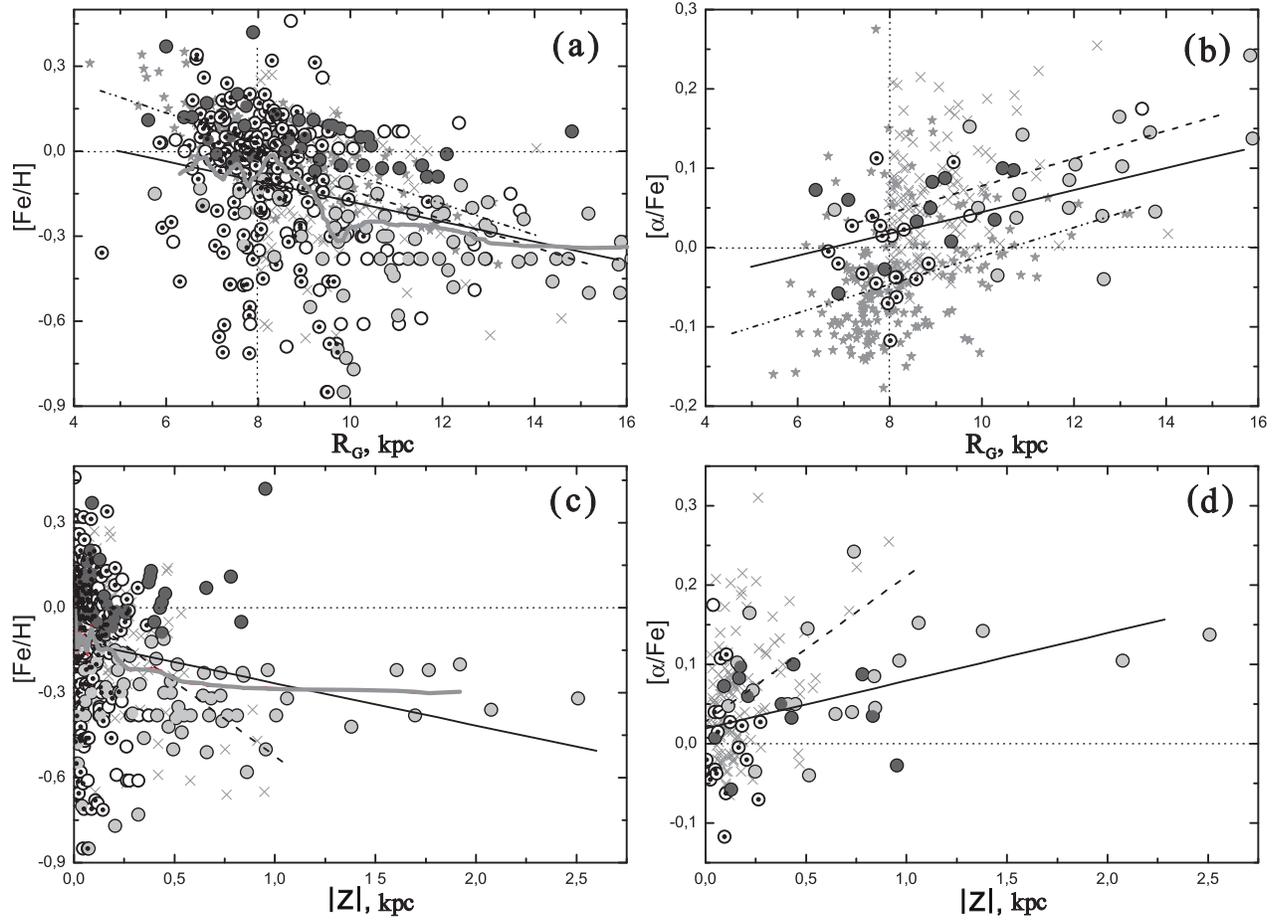}
\caption{Dependences of the metallicity and mean relative 
         abundances of $\alpha$ elements on (a, b) position in the 
         Galaxy and (c, d) distance from the Galactic plane 
         for open clusters (circles), field red giants (x's), 
         and field Cepheids (stars). In view of their
         closeness to the Sun, the apogalactic radii of the 
         orbits and maximum distances of the orbits from the 
         Galactic plane are used for the giants. The thick 
         gray line shows the fit for the clusters, the thin 
         dashed line the fit for the field giants, and the thin
         dotted line the fit for the field Cepheids. The smooth 
         curve is a smoothed trend with a running average for 
         the open clusters. The remaining notation is as in Fig.~1.}
\label{fig4}
\end{figure*}

\newpage

\begin{figure*}
\centering
\includegraphics[angle=0,width=0.50\textwidth,clip]{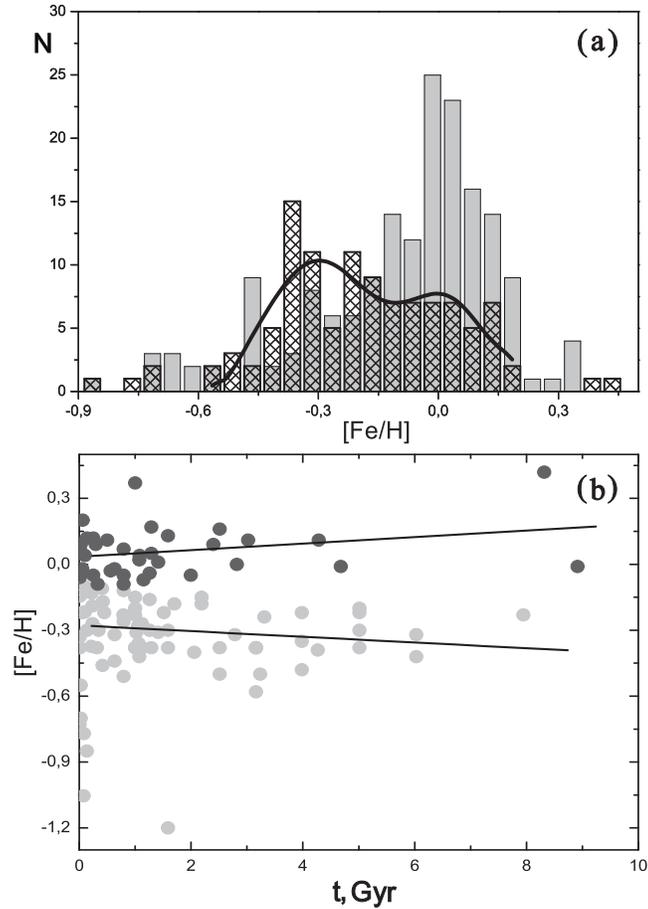}
\caption{(a) Metallicity distributions for open clusters 
         with low, circular orbits (gray) and with high, 
         elongated orbits (hatched). (b) Age-–metallicity 
         dependence for clusters with high, elongated orbits. 
         The curve in (a) is the result of fitting the distribution
         for high, elongated orbits using a sum of two Gaussian curves. 
         The lines in (b) show linear fits for high--metallicity
         (${\rm [Fe/H]} > -0.1$) and low--metallicity clusters. The 
         remaining notation is as in Fig. 1.}
\label{fig5}
\end{figure*}

\newpage


\begin{sidewaystable}
\caption{%
 Statistics of the elemental abundances in the open clusters and their uncertainties}
\bigskip
\label{tproto}
\begin{center}
\begin{tabular}{l|c|c|c|c|c}
\hline \hline
\hspace{-10mm}
& & & & &  \\
\vspace{-6mm}& & & & &  \\

\multicolumn{1}{c|}{\parbox{2.0cm}{Element}}&
\multicolumn{1}{c|}{\parbox{1.9cm}{Number of clusters}}&
\multicolumn{1}{c|}{\parbox{2.0cm}{Mean stated uncertainty}}&
\multicolumn{1}{c|}{\parbox{2.0cm}{Dispersion of the stated uncertainty}}&
\multicolumn{1}{c|}{\parbox{2.7cm}{Number of intersecting determinations}}&
\multicolumn{1}{c}{\parbox{2.7cm}{Dispersion of the calculated mean}}\\
\vspace{-3mm} & & & & &  \\

\hline
O &54 & 0.07 & 0.04 &  56  & 0.09\\
Na&74 & 0.07 & 0.05 &  80  & 0.11\\
Mg&81 & 0.07 & 0.06 &  83  & 0.10\\
Al&71 & 0.07 & 0.07 &  65  & 0.11\\
Si&86 & 0.06 & 0.05 &  100 & 0.10\\
Ca&85 & 0.07 & 0.07 &  109 & 0.09\\
Ti&86 & 0.06 & 0.04 &  93  & 0.09\\
Y &59 & 0.07 & 0.05 &  40  & 0.10\\
Zr&52 & 0.09 & 0.09 &  55  & 0.11\\
Ba&73 & 0.08 & 0.05 &  83  & 0.12\\
La&55 & 0.06 & 0.05 &  41  & 0.11\\
Ce&38 & 0.06 & 0.04 &  12  & 0.04\\
Nd&29 & 0.07 & 0.06 &  9   & 0.09\\
Eu&47 & 0.07 & 0.06 &  26  & 0.12\\

\hline
\end{tabular}
\end{center}
\end{sidewaystable}


\newpage


\begin{sidewaystable}
\caption{%
Elemental abundances and parameters of the Galactic orbits in open clusters
(fragment)}
\bigskip
\label{tproto}
\begin{center}
\begin{tabular}{l|c|c|c|c|c|c|c|c|c|c|c}
\hline \hline
\hspace{-10mm}
& & & & & & & & & & & \\
\vspace{-6mm}& & & & & & & & & & & \\

\multicolumn{1}{c|}{\parbox{0.8cm}{Name}}&
\multicolumn{1}{c|}{\parbox{0.8cm}{$l$, deg}}&
\multicolumn{1}{c|}{\parbox{0.8cm}{$b$, deg}}&
\multicolumn{1}{c|}{\parbox{1.0cm}{$d$, pc}}&
\multicolumn{1}{c|}{\parbox{1.2cm}{$x$, pc}}&
\multicolumn{1}{c|}{\parbox{1.2cm}{$y$, pc}}&
\multicolumn{1}{c|}{\parbox{1.2cm}{$z$, pc}}&
\multicolumn{1}{c|}{\parbox{1.2cm}{$R_{G}$, kpc}}&
\multicolumn{1}{c|}{\parbox{1.2cm}{$V_{R}$, km/s}}&
\multicolumn{1}{c|}{\parbox{1.2cm}{$V_{\Theta}$, km/s}}&
\multicolumn{1}{c|}{\parbox{1.2cm}{$V_{Z}$, km/s}}&
\multicolumn{1}{c|}{\parbox{1.2cm}{$e$}}\\

\vspace{-3mm} & & & & & & & & & & & \\

\hline
NGC 6583&  9.2825& --2.5336&2040&  2011& 329& --90&6.00&-23.8&139.9&  51.5&0.322\\
NGC 6791& 69.9585&  10.9039&5035&  1694&4645&  952&7.89&  8.9&189.7&--34.1&0.115\\
NGC 7789&115.5319& --5.3849&1795& --770&1613&--168&8.92& 16.7&172.3&   1.2&0.222\\
NGC  752&137.1251&--23.2541& 457& --308& 286&--180&8.31&  9.7&213.8&--12.8&0.047\\
NGC 2099&177.6354&   3.0915&1383&--1380&  57&   75&9.38&  0.6&189.6& --1.8&0.150\\

\hline \hline
\hspace{-10mm}
& & & & & & & & & & & \\
\vspace{-6mm}& & & & & & & & & & & \\

\multicolumn{1}{c|}{\parbox{0.8cm}{Name}}&
\multicolumn{1}{c|}{\parbox{0.8cm}{$Z_{max}$, kpc}}&
\multicolumn{1}{c|}{\parbox{1.0cm}{$R_{a}$, kpc}}&
\multicolumn{1}{c}{\parbox{1.2cm}{$R_{p}$, kpc}}&
\multicolumn{1}{c|}{\parbox{1.2cm}{age}}&
\multicolumn{1}{c|}{\parbox{1.2cm}{[Fe/H]}}&
\multicolumn{1}{c|}{\parbox{1.2cm}{Ref. for [Fe/H]}}&
\multicolumn{1}{c|}{\parbox{1.2cm}{[O/Fe]}}&
\multicolumn{1}{c|}{\parbox{1.2cm}{[Na/Fe]}}&
\multicolumn{1}{c|}{\parbox{1.2cm}{[Mg/Fe]}}&
\multicolumn{1}{c|}{\parbox{1.2cm}{[Al/Fe]}}&
\multicolumn{1}{c|}{\parbox{1.2cm}{[Si/Fe]}}\\

\vspace{-3mm} & & & & & & & & & & & \\
\hline

NGC 6583&0.825&6.58&3.37&1.000&  0.37&50&      &    &--0.05&  0.11&  0.01\\
NGC 6791&1.310&8.30&6.58&8.318&  0.42&50&--0.13&0.11&  0.00&--0.06&  0.03\\
NGC 7789&0.172&9.46&6.02&1.413&  0.01&50&  0.13&0.07&  0.18&  0.08&  0.00\\
NGC  752&0.293&8.90&8.11&1.122&--0.02&50&--0.04&0.08&  0.00&  0.08&  0.08\\
NGC 2099&0.082&9.88&7.30&0.347&  0.02&50&  0.15&0.09&  0.27&--0.06&  0.09\\

\hline \hline
\hspace{-10mm}
& & & & & & & & & & & \\
\vspace{-6mm}& & & & & & & & & & & \\

\multicolumn{1}{c|}{\parbox{0.8cm}{Name}}&
\multicolumn{1}{c|}{\parbox{0.8cm}{[Ca/Fe]}}&
\multicolumn{1}{c|}{\parbox{0.8cm}{[Ti/Fe]}}&
\multicolumn{1}{c|}{\parbox{1.0cm}{[Y/Fe]}}&
\multicolumn{1}{c|}{\parbox{1.2cm}{[Zr/Fe]}}&
\multicolumn{1}{c|}{\parbox{1.2cm}{[Ba/Fe]}}&
\multicolumn{1}{c|}{\parbox{1.2cm}{[La/Fe]}}&
\multicolumn{1}{c|}{\parbox{1.2cm}{[Ce/Fe]}}&
\multicolumn{1}{c|}{\parbox{1.2cm}{[Nd/Fe]}}&
\multicolumn{1}{c|}{\parbox{1.2cm}{[Eu/Fe]}}&
\multicolumn{1}{c|}{\parbox{1.2cm}{Ref. for [el/Fe]}}\\

\vspace{-3mm} & & & & & & & & & & & \\
\hline

NGC 6583&--0.01&--0.01&--0.19&    &--0.05&    &    &    &      &67,13\\
NGC 6791&--0.01&  0.01&  0.03&    &--0.05&    &    &    &--0.17&5,17,20,24,47,78\\
NGC 7789&  0.00&--0.04&  0.10&0.12&  0.45&0.11&0.09&0.17&  0.02&61,81,106\\
NGC  752&  0.05&--0.04&--0.01&0.06&  0.38&0.12&0.13&0.16&  0.06&4,22,31,65,88\\
NGC 2099&--0.08&--0.08&--0.07&    &  0.57&0.09&    &0.27&      &81\\

\hline
\end{tabular}
\end{center}
\end{sidewaystable}


\end{document}